\documentclass{emulateapj}
\usepackage{apjfonts}

\def\xmm{\it XMM-Newton\rm}
\def\nh{$N_H$ }

\shorttitle{Revisiting the soft X-ray excess emission in clusters of galaxies}
\shortauthors{J. Nevalainen, M. Bonamente \& J. Kaastra}

\slugcomment{Accepted for The Astrophysical Journal, October 2006}

\begin{document}

\title{Revisiting the soft X-ray excess emission in clusters of galaxies observed with XMM-Newton}

\author{J. Nevalainen}
\affil{Observatory, University of Helsinki, Finland}
\email{jnevalai@astro.helsinki.fi}

\and

\author{M. Bonamente}
\affil{NASA National Space Science Technology Center, Huntsville, AL}
\affil{Department of Physics, University of Alabama, Huntsville, AL}

\and

\author{J. Kaastra}
\affil{SRON, Utrecht, The Netherlands}

\begin{abstract}
We analyze four \xmm\/ galaxy clusters (Abell~1795, Abell~S1101, Abell~1835 and MKW 3s) in order to test 
whether their
soft X-ray excess emission in the 0.2--0.5 keV band as reported by Kaastra et al. (2003) maintains after application of current knowledge
of the \xmm\/  background and calibration.
In this context, we examine the recent claim (Bregman et al., 2006) that the XMM-Newton sub-Galactic H~I column density,  and the accompanying soft excess continuum emission
in the 0.2--0.5 keV band, 
is an artifact of an incorrect background subtraction. We show that since the cluster regions under scrutiny are within 500 kpc of the bright 
cluster center, the X-ray background level is negligible compared to the cluster emission level.
Thus, a ``matched'' background subtraction \'{a} la Bregman et al. (2006) and more traditional background subtraction methods (\'{a} la Kaastra et al. 2003) 
yield the same results for the H~I measurements in the source-dominated central cluster regions. Thus, at least in the central 500 kpc regions of the studied clusters,
the reported soft excess is not a background artifact.
On the other hand, the same simple background-to-cluster flux ratio argument 
allows the possibility that the reported O VII line emission in the 0.5--0.6 keV band 
in the outskirts of some of these clusters (Kaastra et al. 2003)
is due to the geocoronal and heliospheric solar wind charge exchange (Wargelin et al., 2004).
We also study the possibility that the incomplete calibration information in the early phase of the \xmm\/ mission resulted in sub-Galactic H~I column density,
which was interpreted with the presence of a soft excess emitter. 
Our re-analysis of the four \xmm\/ observations with the MOS instruments yields evidence for the soft excess in all clusters.
However, using the PN instrument, the best-fit H~I column densities in Abell~1795, Abell~1835 and MKW 3s are
at odds with the 2003 analysis of Kaastra et al., and in general agreement with
those measured from 21 cm data in the direction of the clusters (Dickey \&
Lockman, 1990). Abell S1101 continues to feature a
sub-galactic NH also with the PN instrument, indicative of soft excess emission in both EPIC detectors.
These differences are compatible with the current level of uncertainty in the calibration of both instruments.

\end{abstract}


\section{Introduction}
The hot phase of the intracluster medium in clusters of galaxies resides in temperatures of T $\sim$ 1--10 keV and thus 
emits strongly X-rays. In addition to this hot gas, there is observational evidence for the existence of a cooler 
(T $\sim$ 0.2-0.5 keV) emission component in several clusters, i.e. the soft excess. 
Using ROSAT PSPC instrument, Bonamente et al. (2002) showed that in a large sample of clusters, about 50\% of the clusters featured 
soft excess in the C band ($\sim$0.25 keV). Using different instruments (\xmm\/ PN and MOS) and methods,
Nevalainen et al. (2003) reported soft excess detection in several clusters in the 0.3--2.0 keV band.
Using \xmm\/,  Kaastra et al. (2003) reported the signature of the cooler component ($\geq$ 20\% excess on top of the hot component emission 
in the 0.2--0.5 keV band) in several clusters. When they analyzed the cluster spectra using only a 
thermal model
for the hot gas, the soft excess emission resulted in sub-Galactic best-fit H~I column densities.
K03 provided further evidence for the cool gas by detecting O VII K$_{\alpha}$ line emission in some clusters
(in the 0.5--0.6 keV band). 

Recent claims by Bregman et al. (2006, B06) indicate
that the soft X-ray excess continuum emission -- or the decrease of the absorption 
column by $\sim 10^{20}$ cm$^{-2}$ compared to 
the 21 cm measurements of Dickey \& Lockman (1990) -- 
in the central 500 kpc regions of clusters Abell 1795, Abell S1101, MKW 3s and 
Abell 1835\footnote{Note that that K03 did not put much confidence in a soft excess in A 1835, because due to 
its large redshift (0.25) that source is very compact and thus the  
energy-dependent point spread function might have affected the results} 
in the 0.2--0.5 keV band, as reported in K03, is  a background subtraction artifact. 
In the following we concentrate on these specific clusters and regions; a 
more general discussion of the soft excess including the cluster outskirts is beyond the scope of this work.

We will examine the effect of the claimed background uncertainties on the measured \nh and the soft excess by 
re-analyzing the XMM-Newton EPIC data of the clusters under scrutiny using 
i) the average blank sky background (Lumb et al, 2002), as in K03
and 
ii) the ``matched'' background which is selected from the blank sky data based on the 
properties of the cluster fields, as in B06.
We will also examine the possibility (as stated in B06) 
that the reported line emission at 0.5--0.6 keV is due to the charge exchange process caused by the solar wind ion collisions 
with the geocoronal and heliospheric neutral atoms. 

We extend the study of the reality of the cluster soft excess in K03 by examining the effect of the improved calibration of \xmm\/ EPIC instruments on 
the soft excess. We carry this out by re-analyzing the K03 clusters using the latest calibration constituents available in March 2006 and 
comparing the results with those published in K03 in Section 5.

\section{X-ray Background}
\label{bkg}
B06 introduced a matched background which is selected from a blank sky data sample based on the values of 
a) the column density as measured from 21 cm data in the direction of the cluster (Dickey \& Lockman, 1990),
b) the ROSAT PSPC 1/4 keV band flux  measured in four points around the cluster at 2 degrees distance and
c) the particle background level of the cluster pointing as measured in the 12--15 keV band.
To emulate the B06 procedure, we use instead the average blank sky data with an appropriate correction (see below).

From their Fig. 4, B06 conclude that the difference between the matched background and the blank sky background 
in case of AS1101 increases rapidly with lower energy. 
While this is true for the absolute values of the difference, both the matched and the blank-sky spectra
increase rapidly at lower energy. Thus the difference between the matched background spectrum and 
the blank sky spectrum is a nearly constant factor of 1.2 in the 0.2--1.0 keV band. 
In the following we therefore estimate the matched background of B06 with the average blank sky spectrum, 
scaled by a constant factor of 1.2.~\footnote{Bonamente et al. (2005) investigated
several ROSAT PSPC observations of the soft X-ray background in the 1/4 keV band, and found a standard
deviation 
of $\sim$30\% among fields within $\sim$10 degrees of 3 clusters, consistent with the $\sim20$ \% background 
increase advocated by B06.}

For the blank sky data we use the data described in detail in N05.
In brief, the N05 procedure consists of \xmm\/ observations selected from a sample of 
blank sky pointings at low \nh ($< 3 \times 10^{20}$ cm$^{-2}$), and double-filtered 
based on the hard band (E $>$ 10 keV for PN, E $>$ 9.5 keV for MOS) and 
soft band (1--5 keV) light curves. The time bins where the hard or soft band count rate deviates from the quiescent level by more than 
$\pm$20\% were rejected, and the accepted data were used to produce co-added event files for each EPIC instrument.

\section{Data processing}
\label{dataproc}
We examine here the XMM-Newton EPIC data of the same clusters A1795, AS1101, A1835 and MKW 3s 
as analysed in K03 and B06.
We use the latest calibration information available in March 2006, i.e. 
revisions up to that of March 6 2006 (see release note XMM-CCF-REL-204 in the XMM-Newton web 
page\footnote{http://xmm.vilspa.esa.es/external/xmm\_sw\_cal/calib/rel\_notes/index.shtml}).

Our procedure follows closely that described in detail in N05. 
We processed the raw data with the SAS version xmmsas\_20050815\_1803-6.5.0 tools epchain and emchain with the default parameters in order to produce the 
event files.
We also generated the simulated out-of-time event file, which we will later use to subtract the events registered during readout of a CCD 
from PN images and spectra.
We filtered the event files with expression ``flag==0'', in order to exclude bad pixels and CCD gaps,
and excluded the regions of bright point sources.
We further filtered the event files including only patterns 0--4 (PN) and 0--12 (MOS).
We used the evselect-3.58.7 tool to extract spectra, images and light curves and the rmfgen-1.53.5 and arfgen-1.66.4 tools
to produce energy redistribution files and the auxiliary response files.

Visual inspection of the images indicates that PN CCDs 10, 11 and 12 in the MKW 3s pointing have anomalously low count rates,
and MOS2 CCD 7 in A1795 pointing has an anomalously high count rate, compared to other CCDs.
Thus we exclude these CCDs from the further analysis.

To reduce the effect of the particle flares, we used the double-filtering method (N05) for screening the data
(see also Section \ref{bkg}).
The resulting exposure times are shown in Table \ref{obsinfo_tab}. 
Note that this is different from what was used in K03 and B06. 
Our method produces cleaner data sets, but since the background does not dominate in these 
bright cluster regions, this difference in the data analysis is expected to make 
a negligible effect on the results.

\section{Background effects on the results} 
\label{dataanal}
Using the data processed above, we examine the emission in the cluster sample using the radial binning scheme from K03
(in order to maintain meaningful statistics) 
but limiting the study to the central 500 kpc as in B06
(in order not to complicate the analysis by additional instrumental problems in the outskirts of the clusters). 
Due to the above choices, we will use radial bins of 
0.0--0.5--1.0--2.0--3.0--4.0--6.0 arcmin for the nearby (z = 0.04--0.06) clusters 
(A1795, AS1101 and MKW 3s) and 0.0--0.5--1.0--2.0 arcmin for the more distant (z = 0.25) A1835 in the spectral analysis.
In this Section, we limit the discussion to PN data, following B06.

\subsection{Background fraction}
\begin{figure}
\plotone{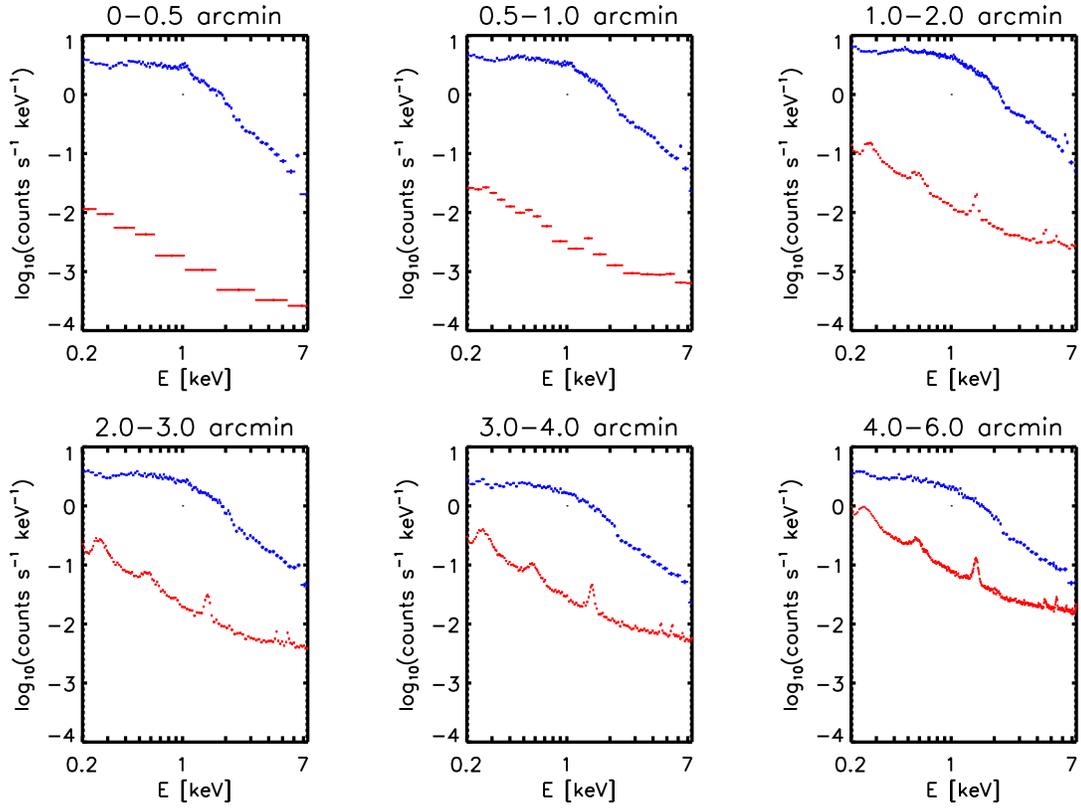}
\caption{The total A1795 emission (upper, blue curves) and the blank sky background from N05 (lower, red curves) observed with PN
\label{sourceandbkg_fig}}
\end{figure}

We evaluate the significance of the background by comparing the total non-background-subtracted cluster spectra with the blank sky spectra 
(see Fig. \ref{sourceandbkg_fig} for A1795).
Using these spectra, we measure the count rates in the 0.2--0.5 keV band, where most of the K03 soft excess is reported.
One can see (Table \ref{bkgtosrc_tab}) that in the central r $\le$ 0.5 arcmin region the background is at 0.1--1\% level of the cluster flux.
Thus, the implied 20\% increase of the B06 background with respect to the N05 blank-sky background 
yields a 0.01--0.1\% decrease in the background-subtracted source fluxes in the central r $\le$ 0.5 arcmin regions.
This effect is 2--3 orders of magnitude too small to affect the observed $\sim$20\% soft excess, 
and the implied background variations cannot explain the soft excess 
phenomenon in the centers of the studied clusters, as will be shown in detail in section \ref{nh}.

In the four clusters analyzed by B06, the highest relative background level, 
40\% of the total emission, is achieved in the 4--6 arcmin region of AS1101 (Table \ref{bkgtosrc_tab}).
The 20\% increase in the background would cause the background-subtracted flux to decrease by 15\%. Only in this region
the B06 background may cause a significant reduction of the soft X-ray fluxes, and an increase of the 
best-fit H~I column density. 

\subsection{Free \nh variation}
\label{nh}
In order to evaluate in detail the effect that the 20\% increase of the background has on the soft excess, 
we performed spectral fits to the cluster data in the 0.2--7.0 keV band.
The modeling of the cluster emission between K03 and B06 is different: while B06 model the gas with an absorbed single MEKAL component,
K03 use an absorbed two temperature VMEKAL model,  where T$_{cool} \equiv 0.5 \ \times $ T$_{hot}$. For detailed description of the treatment of 
different element abundances, see K03. The abundances of both temperature components are forced equal and the  normalizations of the two components are independent.
Also, the radial binning is different between B06 and K03: 
the single temperature modeling of B06 allows smaller regions to be studied with similar statistical accuracy, compared to K03. 
We verified that with the K03 radial binning, both models give consistent values for the \nh.
In the following we will adopt the K03 modeling. \footnote{This modeling allows a factor of 2 variation in the line-of-sight temperatures 
of the hot gas and thus washes out the kind of soft excess (extending up to 2 keV energies) found e.g. in Nevalainen et al., 2003 who used a single 
temperature model for the hot gas in the 2.0--7.0 keV band.} 

\begin{figure}
\plotone{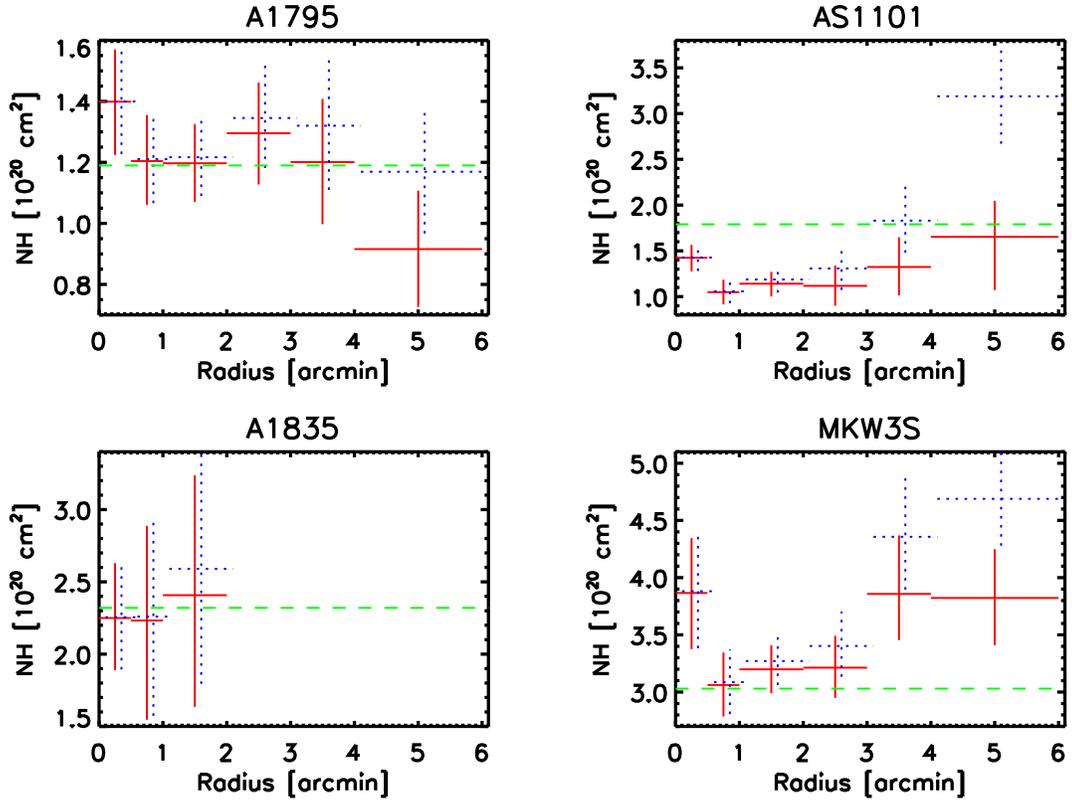}
\caption{The free NH values when using the non-scaled blank sky background from N05 (solid, red lines),
and when using the approximated ``matched'' background of B06 (dotted, blue lines) with 90\% confidence uncertainties.
The B06 radii are shifted by 0.1 arcmin for clarity . The dashed green line shows the Galactig \nh as measured in radio
\label{nhvar_fig}}
\end{figure}

For the background subtraction, we first used 
the N05 double-filtered blank sky data, and then we repeated the analysis increasing the background by 20\% 
in order to approximate the matched background of B06.
In Fig. \ref{nhvar_fig} and Table  \ref{bkgtosrc_tab} we show the results  for the variation in the best-fit \nh
between these two background subtraction methods.
In the central 0.5 arcmin regions the increase in the best-fit \nh 
due to the background variation is 0.001--0.01 $\times 10^{20}$ cm$^{-2}$, 2--3 orders of magnitude smaller than the difference between the B06 and K03 analysis 
(as reported in Fig. 3 of B06).
Only in the 4--6 arcmin regions of AS1101 and MKW 3s the $N_H$ variation is comparable to that reported in B06.

Thus, our analysis of the \xmm\/ PN data of A1795, AS1101, A1835 and MKW 3s clearly indicates that
background uncertainties cannot produce the $N_H$ variations reported in B06.
This is evident from Figure 2, where the N05 background 
is not significant for the majority of the radii of A1795, and from Table 2, in which we show
that the X-ray background contributes to a marginal fraction of the total X-ray counts for the majority
of the regions.

\subsection{Charge exchange}
In addition to the continuum excess in the 0.2--0.5 keV band, K03 also reported a detection of O VII K$_{\alpha}$ line emission 
(i.e. $\sim$20\% enhanchement over the continuum in the 0.5--0.6 keV band) in the outskirts of some of these clusters (AS1101 and MKW 3s),
consistent with the cluster redshift. This would be a direct evidence for the 
presence of cool gas in these clusters. 
B06 argue that this detection is undermined by the charge exchange process:
The process of charge exchange due to solar wind ion collisions between 
neutral heliospheric and geocoronal atoms is shown to produce forbidden O VII K$_{\alpha}$ line 
emission at 561 eV (Snowden et al, 2004 ; Wargelin et al. 2004). This emission varies significantly with time and position on the sky.
Wargelin et al. (2004) showed that the charge-exchange oxygen line may enhance the background by a factor of 2-3 at 0.5--0.6 keV, in time scales of $\sim$months,
similar what B06 shows. 

In K03, the O VII line detections were reported from regions 4--12 arcmin away from the cluster centers. In these regions, the background
at 0.5--0.6 keV becomes substantial (a few 10\%  of the the cluster emission). The high level of background in these regions and the large variation 
observed in Snowden et al. (2004) and Wargelin et al. (2004) can indeed produce effects at the level of 20\% of the cluster emission in the 0.5--0.6 keV band. 
Thus, the simple argument of relative cluster and background fluxes allows the possibility that the O VII line emission is due to the charge exchange.

\section{Calibration}
However, our best-fit $N_H$ values, obtained with the PN data an alysis above, are 
in general agreement with the B06 results (see Table \ref{nh_tab} and Fig. \ref{nhvar_fig}), and inconsistent with 
those reported in the earlier K03 paper.
In the case of AS1101,  the free \nh is still smaller than the value measured from 21-cm data (see also Bonamente et al., 2005).
Thus, ignoring AS1101, our analysis supports the B06 conclusion that the 0.2--0.5 keV band soft excess reported in 
K03 disappears with a re-analysis in the PN data.

Having ruled out the background as a cause for the variance, 
we would like to focus on the point that the 
calibration accuracy, unlike the details of the background, is essential in the analysis of the bright cluster regions. 
The calibration efforts by the XMM-Newton team during past few years have yielded a significant 
improvement in the data processing and energy response 
calculation (see the XMM-CCF release notes in the XMM-Newton web page). 
While we use all the calibration constituents up to that of March 6 2006 (up to and including XMM-CCF-REL-204), B06 use the calibration available in May 2005 (up to 
and including XMM-CCF-REL-189, LLoyd-Davies, {\it priv. comm.}). 
Since there have been no significant published changes in PN calibration between May 2005 and March 2006, 
our PN calibration information is equivalent to that used in B06.  
On the other hand, for the spectral analysis, K03 used the ready-made energy response file {\it epn\_ff20\_sY9\_thin}, 
provided by the \xmm\/ team, available in 2002. 

\begin{figure}
\includegraphics[width=10cm, height=15cm, angle=-90]{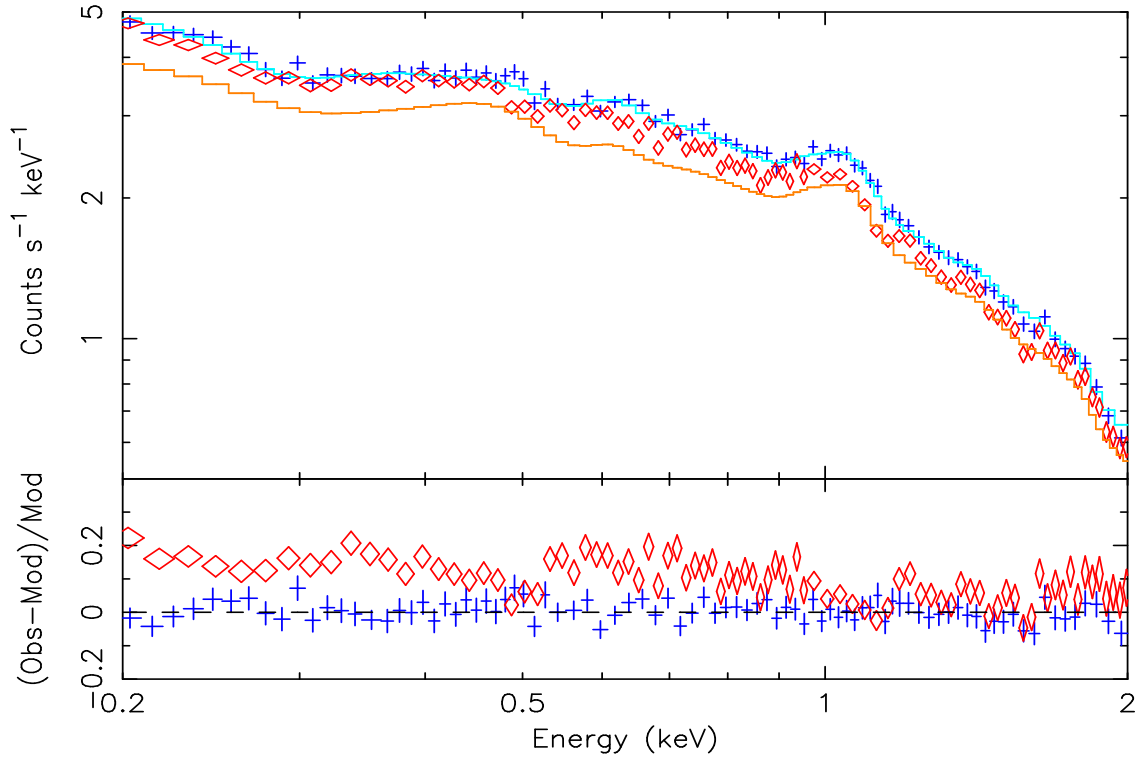}
\caption{{\it upper panel:} PN data of A1795 in the 0.5--1.0 arcmin annulus around the cluster center from K03 
(red diamonds) and N05 (blue crosses). The best 2T fit to N05 data, folded through K03 and N05 responses are shown 
as red and blue lines, respectively. {\it lower panel:} The relative excess of K03 (red diamonds) and N05 (blue crosses) data over the N05 model
\label{K03N05_fig}}
\end{figure}

In order to test whether the change in PN calibration between years 2002 and 2005 is the reason for the different \nh values, 
we compared the K03 spectral analysis of the A1795 0.5--1.0 arcmin annulus data with ours. 
To enable the comparison with K03, we used here only the single pattern events. 
Our best-fit 2-temperature model to the 0.2-7.0 keV band yields \nh consistent with radio measurements and no soft excess. 
Folding this model through K03 responses shows that the data exceeds the folded model count rates by 10--20\% in the 0.2--0.5 keV band 
(see Fig \ref{K03N05_fig}). This difference yields a decrease of 0.5 $\times$ 10$^{20}$ cm$^{2}$ in the best-fit \nh when using K03 responses.
This difference is consistent with that between B06 and K03 as reported in Fig.3 in B06 for A1795 0.5--1.0 arcmin annulus.
This demonstrates clearly that the improvement
in the PN calibration between  October 2002 and August 2005 is the reason for
the difference in  the \nh values between K03 and B06. 

Further indication that the calibration of the instruments is critical for the
determination of the soft excess at low energies is demonstrated by our
separate analysis of the MOS and PN data.
The results of this analysis are shown in Table 3, which shows that
MOS is inconsistent with PN (see Table \ref{nh_tab}) :
while most PN best-fit $N_H$ values are consistent with the 
21-cm measurements (Dickey \& Lockman 1990), the MOS values are significantly lower.
In other words, MOS still exhibits the sub-Galactic \nh and the soft excess, as reported in K03, while PN does not.

\section{Discussion}
We would like to stress that the current results can not be used to reject the 0.2--0.5 keV band soft excess phenomenon for certain, 
because the MOS instrument shows the soft excess in all clusters, and PN provides evidence for the soft excess in AS1101.

Neither can the current results be used to refute the soft excess determined with substantially different analysis methods or
different instruments, such as those of Nevalainen et al. (2003) and Bonamente et al. (2002, 2005).
N03 results could be somewhat affected by the calibration uncertainties we just discovered.
Note however, that N03 used single temperature modeling in the 2-7 keV band and extrapolated the models to 0.5--2.0 keV to find the soft excess, while
B06 and K03 discuss the soft excess in the 0.2--0.5 keV band obtained with a single temperature model (B06) or a 2-temperature model (K03) fit to 
the 0.2--7.0 keV band.
Due to the different modeling the definition of N03 soft excess is different from that in K03 and B06 
and thus the current results are not directly applicable to N03. We plan to address this issue in an upcoming paper.
Bonamente et al. (2002, 2005) use a local background which is not affected by the background subtraction issues discussed in this paper and in B06.

\section{Conclusions}
We re-analyzed the reported (Kaastra et al., 2003) XMM-Newton soft excess continuum detection in the 0.2--0.5 keV band in four clusters of galaxies.
We tested whether the soft excess in the central (r $\le$ 500 kpc) regions of these clusters 
is an artifact produced by incorrect background subtraction (as claimed by Bregman et al., 2006)
or by usage of incomplete calibration information in the early phase of the \xmm\/ mission.
In the above cluster regions, the cluster emission is so strong compared to the background level that 
the details of the background, such as those reported by Bregman et al. (2006) are insignificant.
Thus, the soft excess (or the sub-Galactic $N_H$) measurements remain intact whether we use Kaastra et al. (2003) or Bregman et al. (2006) background method. 
However, the cluster and background flux comparison allows the possibility that the detected O VII line emission 
is due to the heliospheric or geocoronal charge exchange process, rather than due to cool gas. 

Our re-analysis shows that indeed the 0.2--0.5 keV band soft excess reported in Kaastra et al. (2003) disappears
(in all clusters except AS1101) in the latest analysis of the \xmm\/ PN data.
We showed that the difference between the \nh measurements of Bregman et al. (2006) and Kaastra et al. (2003)
is due to changes in the PN calibration between years 2002 and 2005.
The MOS data, on the other hand, still measure 
sub-Galactic $N_H$, and significant amounts of soft excess emission, as in K03.
This provides further evidence that a secure determination of the existence of
the soft excess in \xmm's softest channels (0.2-0.5 keV band), depends
critically on the instrument calibration. 

\acknowledgments
This work is based on observations obtained with XMM-Newton, an ESA science mission with instruments and contributions 
directly funded by ESA member states and the USA (NASA).
JN acknowledges the support from the Academy of Finland. 
SRON is supported financially by NWO, the Netherlands Organization for Scientific Research

\clearpage
  
\clearpage

\clearpage

\clearpage

\begin{deluxetable}{lcccc}
\tablecolumns{5}
\tablecaption{Observation information. t$_{exp}$ is the double-filtered effective exposure time
\label{obsinfo_tab}}
\tablehead{ 
\colhead{name}  & \colhead{obs. ID}   & \colhead{PN t$_{exp}$} & \colhead{MOS1 t$_{exp}$}  & \colhead{MOS2 t$_{exp}$} \\
\colhead{}      & \colhead{}          & \colhead{ks}           & \colhead{ks}              & \colhead{ks}                }
\startdata
A1795    & 0097820101  & 9.8    &  18.8 &   19.8  \\
AS1101   & 0123900101  & 24.8   &  29.7 &   29.7  \\
A1835    & 0147330201  & 5.3    &  6.7  &   7.7  \\
MKW 3s   & 0109930101  & 14.3   &  26.7 &   23.8 \\
\enddata
\end{deluxetable}

\begin{deluxetable}{lcccccccc}
\tablecolumns{9}
\tablecaption{PN results: f$_{bkg}$ shows the count rate ratio of the background to the total cluster observation emission in the 0.2--0.5 keV band. 
$\Delta$ \nh shows the increase in the best-fit \nh, when replacing the non-scaled blank sky background with the approximation to the ``matched'' 
background of B06.
\label{bkgtosrc_tab}}
\tablehead{ 
\colhead{}  & \multicolumn{2}{c}{A1795}       &  \multicolumn{2}{c}{AS1101}     &  \multicolumn{2}{c}{A1835}      &  \multicolumn{2}{c}{MKW 3s}     \\
\colhead{}  & \multicolumn{2}{c}{\hrulefill}  &  \multicolumn{2}{c}{\hrulefill} &  \multicolumn{2}{c}{\hrulefill} &  \multicolumn{2}{c}{\hrulefill} \\
\colhead{r} &  \colhead{f$_{bkg}$} & \colhead{$\Delta$ \nh}  & \colhead{f$_{bkg}$} & \colhead{$\Delta$ \nh}  & \colhead{f$_{bkg}$} & \colhead{$\Delta$ \nh} &  \colhead{f$_{bkg}$} & \colhead{$\Delta$ \nh}          \\
\colhead{arcmin}   &   \colhead{\%}  & \colhead{$10^{20}$ cm$^{2}$} & \colhead{\%}  & \colhead{$10^{20}$ cm$^{2}$} & \colhead{\%}  & \colhead{$10^{20}$ cm$^{2}$} &  \colhead{\%}        & \colhead{$10^{20}$ cm$^{2}$}}   
\startdata
0-0.5    &  0.2  &  0.0008  & 0.3 &  0.003  & 0.4 & 0.006  & 0.7 & 0.02    \\
0.5--1.0 &  0.4  &  0.006   & 0.7 &  0.008  & 2   & 0.03   & 1   & 0.02    \\
1.0--2.0 &  1    &  0.02    & 3   &  0.05   & 9   & 0.2    & 4   & 0.07    \\
2.0--3.0 &  4    &  0.05    & 11  &  0.2    & -   & -      & 8   & 0.2    \\
3.0--4.0 &  8    &  0.1     & 24  &  0.5    & -   & -      & 17  & 0.5    \\
4.0--6.0 &  15   &  0.3     & 42  &  1.5    & -   & -      & 27  & 0.9    \\
\enddata
\end{deluxetable}

\begin{deluxetable}{lccccccccc}
\tablecolumns{10}
\tablecaption{The best-fit parameters and 90\% confidence uncertainties for 2T models in the 0.2--7.0 keV band.
\nh is given in units of  [10$^{20}$ cm$^{-2}$].
K$_{cool}$ and K$_{hot}$ are the VMEKAL normalizations
\label{nh_tab}}
\tablehead{
\colhead{}       & \colhead{}            & \multicolumn{4}{c}{PN}         & \multicolumn{4}{c}{MOS} \\
\colhead{}       & \colhead{}            & \multicolumn{4}{c}{\hrulefill} & \multicolumn{4}{c}{\hrulefill} \\
\colhead{r}      & \colhead{\nh$_{Gal}$}\tablenotemark{a}         &  \colhead{\nh}                   & \colhead{T$_{hot}$} & \colhead{K$_{hot}$}  & \colhead{K$_{cool}$} & \colhead{\nh}        & \colhead{T$_{hot}$} & \colhead{K$_{hot}$} & \colhead{K$_{cool}$} \\
\colhead{arcmin} & \colhead{} &  \colhead{}   & \colhead{keV}       & \colhead{10$^{-3}$}  & \colhead{10$^{-3}$} & \colhead{}  & \colhead{keV}       & \colhead{10$^{-3}$} & \colhead{10$^{-3}$}  }
\startdata
\multicolumn{10}{c}{\bf A1795} \\   
0-0.5    & 1.2  & 1.4$^{+0.2}_{-0.2}$  &  4.5$^{+2.7}_{-0.8}$  & 6.2$^{+2.3}_{-6.0}$   & 2.5$^{+5.9}_{-2.5}$  & 0.9$^{+0.2}_{-0.2}$ & 6.4$^{+1.2}_{-1.8}$   & 3.0$^{+3.9}_{-2.2}$    & 5.2$^{+2.1}_{-3.9}$   \\
0.5--1.0 & 1.2  & 1.2$^{+0.2}_{-0.1}$  &  6.0$^{+1.8}_{-1.1}$  & 6.3$^{+2.9}_{-3.8}$   & 4.4$^{+3.7}_{-2.9}$  & 0.6$^{+0.1}_{-0.1}$ & 6.3$^{+2.7}_{-1.2}$   & 7.2$^{+3.0}_{-5.9}$    & 3.9$^{+5.8}_{-3.2}$   \\
1.0--2.0 & 1.2  & 1.2$^{+0.1}_{-0.1}$  &  7.1$^{+2.1}_{-1.2}$  & 9.4$^{+3.6}_{-5.0}$   & 5.5$^{+5.0}_{-3.7}$  & 0.6$^{+0.1}_{-0.1}$ & 10.7$^{+1.3}_{-0.9}$  & 2.8$^{+5.3}_{-2.8}$    & 13.6$^{+2.9}_{-13.6}$  \\
2.0--3.0 & 1.2  & 1.3$^{+0.2}_{-0.2}$  &  7.8$^{+3.0}_{-7.7}$  & 5.4$^{+3.6}_{-4.7}$   & 4.0$^{+4.6}_{-3.6}$  & 0.5$^{+0.1}_{-0.2}$ & 8.8$^{+3.1}_{-6.1}$   & 6.1$^{+4.1}_{-5.1}$    & 4.8$^{+6.9}_{-4.2}$  \\
3.0--4.0 & 1.2  & 1.2$^{+0.2}_{-0.2}$  &  7.0$^{+5.0}_{-1.6}$  & 4.7$^{+1.9}_{-4.7}$   & 1.8$^{+4.8}_{-1.8}$  & 0.5$^{+0.2}_{-0.2}$ & 9.1$^{+4.2}_{-2.7}$   & 4.1$^{+2.6}_{-4.1}$    & 3.2$^{+4.0}_{-3.2}$   \\
4.0--6.0 & 1.2  & 0.9$^{+0.2}_{-0.2}$  &  7.0$^{+3.2}_{-1.6}$  & 5.3$^{+2.7}_{-4.6}$   & 3.0$^{+4.5}_{-2.8}$  & 0.2$^{+0.2}_{-0.2}$ & 11.4$^{+2.6}_{-5.1}$  & 2.2$^{+6.5}_{-2.2}$    & 6.2$^{+2.4}_{-6.2}$  \\
         &      &                      &                       &                 &               &               &                  &                 &                 \\
\tableline
\multicolumn{10}{c}{\bf AS1101} \\   
0-0.5    & 1.8  & 1.4$^{+0.2}_{-0.1}$   &  4.6$^{+0.3}_{-0.5}$ & 0.2$^{+1.0}_{-0.2}$    & 5.3$^{+0.4}_{-0.9}$  & 0.8$^{+0.2}_{-0.2}$  & 4.8$^{+0.2}_{-0.5}$   & 0.0$^{+1.0}_{-0.0}$    & 5.4$^{+0.1}_{-0.9}$   \\
0.5--1.0 & 1.8  & 1.0$^{+0.2}_{-0.1}$   &  2.7$^{+0.2}_{-0.1}$ & 5.2$^{+0.3}_{-0.6}$    & 0.3$^{+0.6}_{-0.2}$  & 0.7$^{+0.2}_{-0.1}$  & 5.2$^{+0.2}_{-1.2}$   & 0.0$^{+1.5}_{-0.0}$    & 6.0$^{+0.2}_{-1.4}$   \\
1.0--2.0 & 1.8  & 1.1$^{+0.2}_{-0.1}$   &  4.5$^{+0.7}_{-0.6}$ & 1.3$^{+1.2}_{-1.3}$    & 4.8$^{+1.2}_{-1.2}$  & 0.6$^{+0.2}_{-0.1}$  & 5.6$^{+0.1}_{-0.9}$   & 0.0$^{+1.6}_{-0.0}$    & 6.7$^{+0.2}_{-1.5}$   \\
2.0--3.0 & 1.8  & 1.1$^{+0.2}_{-0.2}$   &  4.6$^{+1.0}_{-0.8}$ & 0.5$^{+0.8}_{-0.5}$    & 2.1$^{+0.5}_{-0.8}$  & 0.3$^{+0.2}_{-0.2}$  & 5.4$^{+0.2}_{-1.4}$   & 0.0$^{+1.4}_{-0.0}$    & 3.2$^{+0.1}_{-1.3}$   \\
3.0--4.0 & 1.8  & 1.3$^{+0.3}_{-0.3}$   &  4.1$^{+1.0}_{-1.0}$ & 0.4$^{+0.7}_{-0.4}$    & 1.2$^{+0.4}_{-1.2}$  & 0.3$^{+0.3}_{-0.3}$  & 4.7$^{+0.6}_{-2.6}$   & 0.2$^{+1.7}_{-0.2}$    & 1.7$^{+0.3}_{-1.7}$   \\
4.0--6.0 & 1.8  & 1.7$^{+0.4}_{-0.6}$   &  3.5$^{+0.6}_{-0.6}$ & 0.3$^{+0.6}_{-0.3}$    & 1.6$^{+0.3}_{-0.6}$  & 0.0$^{+0.3}_{-0.0}$  & 4.5$^{+0.7}_{-1.1}$   & 0.2$^{+0.9}_{-0.2}$    & 1.9$^{+0.3}_{-0.8}$   \\
         &      &                 &                  &                 &                &               &                 &                  &                 \\
\tableline
\multicolumn{10}{c}{\bf A1835} \\   
0-0.5    & 2.3  & 2.2$^{+0.4}_{-0.3}$   &  7.8$^{+4.6}_{-2.4}$  & 4.7$^{+3.4}_{-4.7}$   & 3.2$^{+4.9}_{-3.2}$  & 2.3$^{+0.5}_{-0.6}$ & 9.8$^{+2.9}_{-0.9}$   & 2.2$^{+7.1}_{-2.2}$   & 5.2$^{+2.5}_{-5.2}$  \\
0.5--1.0 & 2.3  & 2.2$^{+0.7}_{-0.7}$   &  15.0$^{+3.5}_{-7.6}$ & 0.0$^{+4.2}_{-0.0}$   & 4.2$^{+0.2}_{-4.2}$  & 1.1$^{+1.3}_{-1.1}$ & 12.1$^{+11.4}_{-4.8}$ & 2.3$^{+1.8}_{-2.3}$   & 1.3$^{+2.7}_{-1.3}$   \\
1.0--2.0 & 2.3  & 2.4$^{+0.8}_{-0.8}$   &  9.7$^{+10.1}_{-3.4}$ & 2.4$^{+1.4}_{-2.4}$   & 1.0$^{+2.8}_{-1.0}$  & 1.9$^{+1.2}_{-1.1}$ & 14.8$^{+10.7}_{-7.1}$ & 1.7$^{+2.5}_{-1.7}$   & 2.2$^{+2.0}_{-2.2}$   \\
         &      &                 &                  &                 &                &               &                 &                  &                 \\
\tableline
\multicolumn{10}{c}{\bf MKW 3s} \\   
0-0.5    & 3.0  & 3.9$^{+0.5}_{-0.5}$   &  4.8$^{+2.0}_{-1.9}$   & 1.1$^{+1.7}_{-1.1}$   & 1.6$^{+1.1}_{-1.6}$  & 2.9$^{+0.5}_{-0.5}$ & 3.2$^{+0.3}_{-0.2}$   & 3.1$^{+0.1}_{-0.2}$    & 0.0$^{+0.2}_{-0.0}$   \\
0.5--1.0 & 3.0  & 3.1$^{+0.3}_{-0.3}$   &  5.2$^{+1.7}_{-1.9}$   & 1.5$^{+2.5}_{-1.5}$   & 2.5$^{+1.5}_{-2.5}$  & 2.8$^{+0.2}_{-0.2}$ & 3.5$^{+0.3}_{-0.1}$   & 4.8$^{+0.1}_{-0.4}$    & 0.0$^{+0.4}_{-0.0}$   \\
1.0--2.0 & 3.0  & 3.2$^{+0.2}_{-0.2}$   &  4.9$^{+2.1}_{-2.5}$   & 3.8$^{+3.5}_{-3.8}$   & 3.7$^{+3.6}_{-3.7}$  & 2.9$^{+0.2}_{-0.2}$ & 3.6$^{+0.3}_{-0.1}$   & 8.1$^{+0.1}_{-0.8}$    & 0.0$^{+0.8}_{-0.0}$   \\
2.0--3.0 & 3.0  & 3.2$^{+0.3}_{-0.3}$   &  4.8$^{+1.8}_{-4.7}$   & 2.6$^{+1.6}_{-2.0}$   & 1.9$^{+1.9}_{-1.7}$  & 2.6$^{+0.2}_{-0.2}$ & 4.1$^{+3.3}_{-0.4}$   & 5.2$^{+0.7}_{-5.2}$    & 0.6$^{+4.3}_{-0.6}$   \\
3.0--4.0 & 3.0  & 3.9$^{+0.5}_{-0.4}$   &  4.3$^{+1.3}_{-1.1}$   & 1.6$^{+1.1}_{-1.4}$   & 1.3$^{+1.4}_{-1.2}$  & 2.4$^{+0.3}_{-0.3}$ & 7.0$^{+0.4}_{-3.2}$   & 0.0$^{+3.3}_{-0.0}$    & 3.9$^{+0.1}_{-3.9}$   \\
4.0--6.0 & 3.0  & 3.8$^{+0.4}_{-0.4}$   &  3.3$^{+0.7}_{-0.5}$   & 2.2$^{+1.0}_{-1.1}$   & 1.9$^{+5.2}_{-1.0}$  & 2.1$^{+0.3}_{-0.2}$ & 4.4$^{+1.1}_{-1.2}$   & 2.1$^{+2.2}_{-1.6}$    & 3.0$^{+1.6}_{-2.3}$   \\
         &      &                 &                  &                 &                &               &                 &                  &                 \\
\enddata
\tablenotetext{a}{Galactic column density [10$^{20}$ cm$^{-2}$] measured in radio (Dickey \& Lockman 1990)}
\end{deluxetable}


\begin{thebibliography}{}
\bibitem[]{} Bonamente, M., Lieu, R.,  Mittaz, J., et al., 2005, ApJ, 629, 192
\bibitem[]{} Bonamente, M., Lieu, R.,  Joy, M., et al., 2002, ApJ, 576, 688
\bibitem[]{} Bregman, J., \& Lloyd-Davies, E., 2006 (B06), ApJ in press, astro-ph/0602527 
\bibitem[]{} Dickey, J. \& Lockman, F., 1990, ARAA, 28, 215.
\bibitem[]{} Kaastra, J., Lieu, R., Tamura, R., et al., 2003 (K03), A\&A, 397, 445,  
\bibitem[]{} Lumb, D., Warwick, R., Page, M., \& De Luca, A., 2002, A\&A, 389, 93
\bibitem[]{} Markevitch, M.,  Bautz, M., Biller, B., et al., 2003, ApJ, 583, 70
\bibitem[]{} Nevalainen, J., Lieu, R., Bonamente, M, \& Lumb, D., 2003, ApJ, 584, 716
\bibitem[]{} Nevalainen, J., Markevitch, M., \& Lumb, D., 2005 (N05), ApJ, 629, 192
\bibitem[]{} Snowden, S., Collier, M., \& Kuntz, K., 2004, ApJ, 610, 1182
\bibitem[]{} Wargelin, B., Markevitch, M., Juda, M., et al., 2004, ApJ, 607, 596
\end{thebibliography}
\end{document}